\documentclass[12pt]{article}
\usepackage{epsfig}
\usepackage{amsfonts}
\usepackage{amscd}
\usepackage{latexsym}
\usepackage{amsmath,amssymb}
\usepackage{verbatim}
\usepackage{setspace}
\usepackage{color}
\usepackage{cite}
\usepackage[textheight=9in, textwidth=6.5in, letterpaper]{geometry}

\usepackage{hyperref}
\hypersetup{
    colorlinks,
    citecolor=black,
    filecolor=black,
    linkcolor=black,
    urlcolor=black
    linktoc=all
}

\usepackage{graphicx}
\usepackage{tikz}
\usepackage{tikz-3dplot}
\usetikzlibrary{calc}
\usetikzlibrary{positioning}
\usepackage{float}
\usepackage{subfig}
\usepackage[T1]{fontenc}
\usepackage{amsmath, amssymb, amsthm, fullpage}
\usepackage{doc}
\usepackage{ifthen, color, fancyvrb}
\usepackage{nextpage}

\usepackage{wrapfig}
\usepackage{parskip}
\usepackage{authblk}
\usepackage[makeroom]{cancel}

\def\be{\begin{equation}}
\def\ee{\end{equation}}
\def\bz{{\bar z}}
\def\p{\partial}
\def\ci{{\cal I}}
\def\ip{${\cal I}^+$}
\def\gzz{\gamma_{z\bz}}
\def\be{\begin{equation}}
\def\ee{\end{equation}}
\def\bea{\begin{eqnarray}}
\def\eea{\end{eqnarray}}
\def\<{\langle }
\def\>{\rangle}
\def\cz{{{\Sigma}}}\def\cf{\Sigma_F}
\def\v{\epsilon}
\def\ig{{{\cal I}^+_>}}
\def\tr{{\rm tr}}
\begin{document}
\begin{titlepage}
\unitlength = 1mm

\ \\

\vskip 3cm

\begin{center}

{ \LARGE {\textsc{Area, Entanglement Entropy and Supertranslations at Null Infinity}}}

\vspace{0.8cm}
Daniel Kapec, Ana-Maria Raclariu and Andrew Strominger

\vspace{1cm}

{\it  Center for the Fundamental Laws of Nature, Harvard University,\\
Cambridge, MA 02138, USA}\\

\begin{abstract}
 The area of a cross-sectional cut $\Sigma$ of future null infinity (\ip) is infinite. We define a finite, renormalized 
 area by subtracting the area of the same cut in any one of the infinite number of BMS-degenerate classical vacua. The renormalized area acquires an anomalous dependence on the choice of vacuum. 
We relate it to the modular energy, including a soft graviton contribution, of the region of \ip\ to the future of $\Sigma$. Under supertranslations, the renormalized area shifts by the supertranslation charge of $\Sigma$.  In quantum gravity, we conjecture a bound relating the renormalized area
to the entanglement  entropy across $\Sigma$ of the outgoing quantum state on \ip. 
   \end{abstract}

\vspace{1.0cm}

\end{center}

\end{titlepage}

\pagestyle{empty}
\pagestyle{plain}

\def\vx{{\vec x}}
\def\p{\partial}
\def\po{$\cal P_O$}

\pagenumbering{arabic}

\tableofcontents

\section{Introduction}
The Bekenstein-Hawking area-entropy law \cite{Bekenstein:1972tm,Hawking:1974sw}
\be \label{dalaw} S_{BH}={{\rm{Area}}\over 4\hbar G} \ee
ascribes an entropy to a null surface proportional to its cross-sectional area in Planck units. This law has 
a number of fascinating generalizations \cite{Unruh:1976db,Gibbons:1977mu,Gibbons:1976ue, Wald:1993nt,Iyer:1994ys,'tHooft:1993gx,Fiola:1994ir,Susskind:1994vu,Jacobson:1995ab,Strominger:1996sh,Maldacena:1997re,Witten:1998qj,Susskind:1998dq,Gubser:1998bc,Hawking:2000da,Fischler:1998st,Strominger:2003br, Ryu:2006bv,Ryu:2006ef,Casini:2008cr}, including the Bousso bound \cite{Bousso:1999xy,Bousso:1999cb,Bousso:2002ju,Flanagan:1999jp,Bousso:2014sda,Bousso:2014uxa,Bousso:2015mna  } which bounds  the change in the area to the 
entropy flux through the null surface. 

One of the most interesting null surfaces is future null infinity (\ip) which is a future boundary of asymptotically Minkowskian spaces. It is a universal observer horizon for all eternal observers which do not fall into black holes. It is natural to try to relate the change in the area of cross-sectional `cuts' $\Sigma$ of \ip\ to the energy or entropy flux across \ip. An immediate obstacle is that both the areas and area changes of such cuts are infinite. The Bousso bound is obeyed but in a trivial manner. 

In this paper we define a finite  renormalized area of cuts of \ip\ and conjecture a nontrivial bound relating it to  the entropy radiated through \ip.   A family of regulated null surfaces parametrized by $r$ which approach \ip\ for $r\to \infty$ is introduced. 
For finite $r$ these have finite area for any cut $\Sigma$. We then define a subtracted area by subtracting the area of the same cut in a fiducial vacuum geometry.  The gravitational vacuum has an infinite degeneracy labeled by an arbitrary function $C_0$ on the sphere at \ip \cite{Strominger:2013jfa}.\footnote{Prescient early discussions of vacuum degeneracy are in  \cite{ash,ashb,Balachandran:2014hra}.}
 Under BMS supertranslations, also parameterized by an arbitrary function (denoted $f$) on the sphere,  $C_0\to C_0+f$ and these vacua transform into one another. We show that the subtracted area, denoted $A^\Sigma_0$, is finite (and typically negative) in the $r\to \infty$ limit. However it retains `anomalous' dependence on the choice of a fiducial $C_0$. 

This renormalized area $ A^\Sigma_0$ is found to have several interesting properties. When $C_0$ coincides with the physical vacuum at the location of the cut, $ A^\Sigma_0$ is the negative of the so-called modular energy of the region $\ci^+_>$ lying to the future of $\Sigma$, including `soft graviton' terms which are linear in the Bondi news. It tends to increase towards the far future, and asymptotically reaches zero from below when $C_0$ coincides with the asymptotic future vacuum. Moreover,  under supertranslations it shifts by the supertranslation charge on $\Sigma$. 

In quantum gravity, the outgoing quantum state is supported on \ip. The cut $\Sigma$ divides \ip\ into two regions, and a quantum entanglement entropy $S^{\rm ent}_0$ of the portions of the outgoing quantum state on opposing sides of the cut is expected. In principle, unlike the entanglement across generic fluctuating interior surfaces, $S^{\rm ent}_0$ should be well defined because gravity is weakly coupled near the boundary. However it is beset by both ultraviolet (UV) divergences from short wavelength entanglements near the cut and infrared (IR) divergences from soft gravitons. A choice of vacuum is required for subtraction of UV divergences, so $S^{\rm ent}_0$ will also acquire an `anomalous' dependence on the fiducial $C_0$.  We will not try to give a precise definition of $S^{\rm ent}_0$ herein which would, among other things, require a decomposition of the soft graviton Hilbert space.\footnote{A relevant discussion appears in \cite{Casini:2013rba,Donnelly:2014fua,Donnelly:2015hxa}. } Nevertheless we will motivate a conjecture that a suitably defined $S^{\rm ent}_0$ obeys the bound 
\be -{A^\Sigma_0\over 4\hbar G} \ge S^{\rm ent}_0(\Sigma)\ee
for any cut $\Sigma$ of \ip. Typically, at late times both sides of this equation are positive and decreasing. This relation incorporates the BMS structure at \ip\ into the study of the relation between area and entanglement entropy. 

Our results are plausibly relevant to and were motivated by the black hole information paradox. 
A unitary resolution of this paradox would amount, roughly speaking,  to showing  that late and early time Hawking emissions are correlated in such a way that, for a pure incoming state,  the full quantum state on \ip\ is a pure state. However a more precise BMS-invariant statement is needed. One would like to compute the entanglement entropy $S^{\rm ent}_0$ across any cut $\Sigma$. Naively, one expects that 
it approaches zero for all cuts in the far past and far future and has a maximum somewhere in the middle, possibly at the Page time \cite{Page:1993wv,Page:2013dx}. Given both the IR and UV subtractions needed to define $S^{\rm ent}_0$, the resulting anomalies in supertranslation invariance and the discovery of soft hair \cite{Hawking:2016msc,Strominger:2014pwa},  it is not obvious to us what precisely the expectation following from unitarity should be. In particular, the requirement that $S^{\rm ent}_0$ vanish in the far future is not  fully supertranslation invariant. We do not attempt to resolve these issues herein. Rather we view the present effort as a first step in obtaining a precise statement of the black hole information paradox.

This paper is organized as follows. Section 2 contains preliminaries and notation. In section 3 we define a renormalized area $ A^\Sigma_F$ in which we subtract the area associated to  the asymptotic future vacuum and relate it to the `hard' modular energy of the region to the future of the cut. In section 4 we show that $A^\Sigma_F$ varies under supertranslations into the hard part of the supertranslation charge. Section 
5 introduces the more general renormalized $A^\Sigma_0$ involving an arbitrary vacuum subtraction. Its variation under supertranslations is shown to involve the full modular energy including soft graviton contributions. 
In section 6 we motivate and conjecture a bound relating the renormalized area to the entanglement entropy which can be viewed as the second law of \ip. 

Throughout this paper we assume for simplicity that the geometry reverts to a vacuum in the far future and that all flux though \ip\ is gravitational.  This highlights many of the salient features but a  treatment of more general cases would be of interest. 
\section{Preliminaries}
In retarded Bondi coordinates, asymptotically flat  metrics \cite{Bondi:1962px,Sachs:1962wk,Sachs:1962zza,Ashtekar:2014zsa} near \ip\   take the form 
\be\label{met} \begin{split}
ds^2=&-du^2 -2dudr + 2r^2 \gamma_{z\bz}dzd\bz \\
&+\frac{2m_B}{r}du^2+ rC_{zz}dz^2 +rC_{\bz\bz}d\bz^2 + D^zC_{zz}dudz + D^\bz C_{\bz\bz}dud\bz + \dots   
\end{split}
\ee
Here $\gzz$ is the round metric on the unit $S^2$  and 
$D_z$ is the associated covariant derivative. 
Defining \bea \label{cft} N_{zz}&=&\p_uC_{zz},\cr T_{uu}&=&\frac12 N^{zz}N_{zz}, \cr
U_z&=&iD^zC_{zz}, ~~~~U=U_zdz+U_\bz d\bz,\cr
V_z&=&iD^zN_{zz},~~~~V=V_zdz+V_\bz d\bz,\cr
 \v&=& i\gzz dz\wedge d\bz,  \eea
the leading order vacuum constraint equation reads 
\be\label{cst}
 \p_u m_B\;du\wedge \v=-\frac{1}{2}T_{uu}\; du\wedge \epsilon -\frac14du \wedge dV .   
\ee
We could easily add a matter contribution to $T_{uu}$ but we omit this for brevity. 
We assume that near the future boundary $\ci^+_+$ of \ip\ 
the spacetime reverts to a vacuum so that 
\be
 m_B|_{\ci^+_+}=0,~~~C_{zz}|_{\ci^+_+}=-2 D_z^2C_F,  
  \ee 
for some function $C_F(z,\bz)$. In the quantum theory we denote the corresponding vacuum state by $|C_F\>$.
Given $C_F$ and the Bondi news tensor $N_{zz}$,  the mass aspect $m_B$ is determined by integrating the constraint equation (\ref{cst}) backwards from $\ci^+_+$. 

 Asymptotically flat spacetimes admit an infinite dimensional symmetry group, known as the Bondi-Metzner-Sachs (BMS) group \cite{Bondi:1962px,Sachs:1962wk,Sachs:1962zza}. The supertranslations are labeled by an arbitrary function $f(z,\bar{z})$ on the $S^2$ and are generated by the vector fields 
 \be \label{stn} 
 \xi= f\p_u -{1\over r}(D^z f \p_z + D^\bz f\p_\bz) 
+\frac12D^2 f\p_r,    ~~~~~D^2=2D^zD_z.  
\ee 
Infinitesimal supertranslations act on the geometry as \cite{Barnich:2010eb,Barnich:2011mi} 
\begin{eqnarray} \label{str} \delta_fC_{zz}&=&fN_{zz}-2D_z^2f,\cr\delta_f C_F&=&f,\cr
\delta_fU_z&=&fV_z+ iD^zfN_{zz}-i D^2D_zf, \hspace{5pt} \cr
        \delta_f m_B&=&f\p_um_B+\frac14 D_z^2fN^{zz}+\frac14 D_\bz^2fN^{\bz\bz}+\frac{i}{2}\p_zfV^z-\frac{i}{2}\p_\bz f V^\bz,\cr \delta_f T_{uu}&=&f\p_uT_{uu}.   \end{eqnarray}
        
\section{Renormalizing the area }

We wish to study the area of a cut $\Sigma$ of \ip\ defined by 
\be u=u_\Sigma(z,\bar{z} ) \ee in the geometry (\ref{met}) and also to study its variation under supertranslations of $\Sigma$
\be u_\Sigma \to u_\Sigma+f\ee  with the geometry  held fixed. Of course this area is infinite so we must introduce both a regulator and a subtraction. 
We regulate the area by the replacement of \ip\ with the past lightcone of  a point which approaches $i^+$. 
For  the flat Minkowski metric the null hypersurface 
\be 
r=-\frac12 (u- u_0)   
 \ee
approaches \ip\ for $u_0\to \infty$ with $u$ held fixed. 
More generally we solve the ODE
\begin{equation}\label{ode}
\left(1 - \frac{2m_B}{r}\right) du +2 dr = 0,   
\end{equation}
which guarantees that the surface is null, and choose the  integration constants at each $(z,\bz)$ so that the surface lies at large radius, approaching infinity,  for finite $u$.\footnote{The generic such surface  
will terminate at a cusp rather than a point, but this will not matter as the quantities considered below do not have contributions from the endpoint of the surface.} The null condition (\ref{ode}) has ${1 \over r}$  corrections. For brevity such corrections are suppressed  here and hereafter whenever they drop out of the large-$r$ limit. 
Equation (\ref{ode}) can be rewritten as
\begin{equation}
\label{r-eq}
\frac{dr^{2}}{du} = 2m_B - r .    
\end{equation}
The area of a cut $\cz$ of \ip\ defined by $u = u_{\Sigma}(z, \bar{z})$  then follows from the metric induced from (\ref{met}) and is given by 
\begin{equation}\label{az}
A(\cz, N_{zz}, C_F ) = \int_\cz d^2z \sqrt{\det g} =\int_\cz \left( r^{2}\v - \frac{1}{2}du_{\Sigma}\wedge U\right)   .
\end{equation}

Both the area {(\ref{az}) as well as its variation with respect to retarded time are divergent in the large-$r$ limit of interest. 
A subtraction  is necessary to obtain a finite result. 
We define a fiducial  `$C_F$-vacuum' in which the news $N_{zz}$ vanishes and $C_{zz}=-2D_z^2C_F$ on all of \ip.
This flat geometry coincides with (\ref{met}) at late times. 
A fiducial null hypersurface in this fiducial spacetime solving (\ref{r-eq}) (with $m_B=0$) can then be found which coincides exactly with the solution of (\ref{ode}) in (\ref{met}) at late times.  A subtracted area, with a finite large-$r$ limit, may then be obtained by subtracting the area of the fiducial hypersurface: %
\begin{equation}\label{azr}\begin{split}A^\Sigma_F &= A(\cz, N_{zz}, C_F )-A(\cz, 0, C_F ) \\
&= \int_\Sigma \left[(r^{2} - r_{0}^{2})\v + \frac{1}{2} du_{\Sigma}\wedge \Delta U\right].   
\end{split}
\end{equation}
Here $\Delta U=U_F-U_\Sigma $ is the change in $U $ and $r_0$ is the radius in the fiducial vacuum. Using (\ref{r-eq}), we have
\begin{equation}
\frac{d}{du}(r^{2} - r_{0}^{2}) = 2m_B.  
\end{equation}
Integrating this equation from $\ci^+_+$ to $u_{\Sigma}$, one finds
\begin{equation}
r^{2}(u_{\Sigma},z,\bz) - r_{0}^{2}(u_{\Sigma},z,\bz)  = -\int_{u_{\Sigma}}^{\infty}du\;2 m_B (u,z,\bz) .   
\end{equation}
The finite renormalized area is then given by
\begin{equation}\label{daone} \begin{split}
A^\Sigma_F &= 
 -\int_\ig du\wedge \left[2m_B\v - \frac{1}{2}du_{\Sigma}\wedge  V\right]  , 
\end{split}
\end{equation}
where $\ig$ is the three-dimensional region of \ip\ lying to the future of the cut $\Sigma$. 
Using the constraint equation and  the identity 
\be \int_{\ci^+_>}du\wedge du_\Sigma \wedge V= \int_{\ci^+_>}(u-u_\Sigma) du \wedge d V =- 
 \int_{\Sigma} u_\Sigma d \Delta U,  \ee
 the renormalized area can be rewritten 
\be\label{datwo}
 A_F^\Sigma= -\int_\ig  (u-u_{\Sigma}) T_{uu}du \wedge\v=-\int_\cz d^{2}z \gamma_{z\bar{z}}\int_{u_{\Sigma}}^{\infty} (u-u_{\Sigma})T_{uu}du.   
\ee
We refer to this as the (negative of the) hard modular energy of the region $\ci^+_>$. We note that $A_F^\Sigma$ is typically negative and increases to zero in the far future due to the subtraction scheme. 
\section{Supertranslations} $ A^\Sigma_F$ is strictly invariant under coordinate transformations which both move the cut and transform the physical and subtraction geometries. In particular, $A^\Sigma_F$ is invariant if we simultaneously shift the cut $u_\Sigma \to u_\Sigma + f$ and supertranslate the geometry by the inverse transformation.  However, one can consider evaluating the subtracted area on a supertranslated cut, sending $u_{\Sigma} \rightarrow u_{\Sigma} + f$ while keeping the geometry fixed. Starting from either  (\ref{daone}) or (\ref{datwo}) one easily finds 
\begin{equation}\label{atr}\delta_{f} A_F^\Sigma
=\int_\cz  f\left[2m_B\v- \frac{1}{2}d\Delta U  \right]= \int_\ig fT_{uu} \; du\wedge \v      .
\end{equation}
The right hand side is the hard part of the supertranslation charge on $\ig$. 
Alternately, (\ref{atr}) may be derived by infinitesimally supertranslating the geometry according to (\ref{str}). 

\section{A general subtraction}
The subtraction used in (\ref{azr}) to obtain a finite area change has a teleological nature: we must know which vacuum the geometry settles into in the far future in order to define $ A^\Sigma_F$. In this section we consider a more general, non-teleological subtraction of the area of  $\Sigma$ at $u=u_\Sigma$ in the null hypersurface defined by solving (\ref{r-eq}) in an
 arbitrary vacuum characterized by the arbitrary function $C_0$ with  $C_{0zz}=-2D_z^2C_0$.  Unlike the case in (\ref{azr}),
the subtracted geometry is not identical to the physical one at late times, and so the late-time contributions are not manifestly finite or well-defined. To characterize the resulting  ambiguity  we introduce a late-time cutoff by terminating both surfaces at a final cut $\cf$ at $u=u_F(z,\bz)$, in the late-time vacuum region with $m_B=0$.\footnote{It would be interesting to analyze the more generic case of the area change over a more general finite interval.}  One finds
\be 
A^\Sigma_0 = A^\Sigma_F +\frac12 \int_\cz d(u_{\Sigma}-u_F)\wedge  (U_{0}-U_{F})   ,
\ee 
where $U_0$ and $U_F$ are constructed from $C_0$ and $C_F$ according to (\ref{cft}). 
As may be easily verified, this expression is invariant if we supertranslate the physical geometry, the fiducial vacuum $C_0$ and both cuts at $\cz$ and $\cf$. 
We now restrict consideration to the case $u_F = $ constant, in which case this expression reduces to 
\be
A^\Sigma_0 =  -\int_\ig du \wedge (u-u_{\Sigma})(\v T_{uu}+\frac12 dV)+\frac12 \int_\cz du_{\Sigma}\wedge  (U_{0}-U) .
\ee

Fixing the geometry and varying $u_\Sigma \to u_\Sigma +f$, we find
\be
\delta_f A^\Sigma_0=\int_{\Sigma} f\left[	2  m_B\epsilon -		\frac{1}{2}d(U_0-U)	\right]  .
\ee
The right hand side is the full supertranslation charge in the special case $U=U_0$ on $\Sigma$. 

\section{An area-entropy bound conjecture}

Given a cut $\cz$ and  a vacuum state $|C_0\>$ on all of $\ci^+$ we may define a density matrix on the region $\ci^+_>$ to the future of $\cz$ 
by 
\be \sigma_0=\tr_< |C_0\> \< C_0|,\ee
where the trace is over the region prior to $\cz$, and the dependence on 
the choice of cut is suppressed. 
Similarly for an excited state $|\Psi\>$ we define the density matrix on $\ci^+_>$
\be \rho=\tr_<|\Psi \>\<\Psi |.\ee 
We normalize so that $\tr\rho=\tr\sigma_0=1.$ 
The modular hamiltonian which measures local Rindler energies relative to $|C_0\>$ is 
\be - \ln \sigma_0.\ee
$\sigma_0$ has contributions from the entanglement of both hard and soft modes across the surface $\Sigma$. Hard mode entanglements contribute \cite{Wall:2010cj,Wall:2011hj}\footnote{Note that our normalization of the stress energy tensor as defined in (\ref{cst}) differs by a factor of $8\pi G$ from some other references.}
\be
 -\ln \sigma_0|_{hard}= \frac{1}{4\hbar  G} \int_\ig du \wedge  (u-u_{\Sigma})\v \hat T_{uu}+{\rm constant}=-\frac{\hat{A}_F^\Sigma}{ 4\hbar G}+{\rm constant},
\ee
where here $\hat T_{uu}$ and $\hat A$ are both operators and the constants depend on the normal ordering prescription.
It would be extremely interesting, but beyond the scope of this paper, to regulate, define and compute the soft contributions to $\sigma_0$. The precise form of $\sigma_0$ may well depend on the renormalization scheme. Here we simply conjecture, motivated by the structures encountered in the previous section, that these contributions can be defined in such a way
that 
 \be\label{cnj} -\ln \sigma_0= - \frac{\hat{A}^\Sigma_0}{ 4\hbar G}+{\rm constant},\ee
where the operator-valued area appearing here is 
\be
\hat A^\Sigma_0 =  -\int_\ig du \wedge (u-u_{\Sigma})(\v \hat T_{uu}+\frac12 d\hat V)+\frac12 \int_\cz du_{\Sigma}\wedge  (U_{0}-\hat U).
\ee 
We interpret the  first term as the full modular Hamiltonian, including soft terms. The last is a soft term which vanishes when $U_0=\hat U$ on the cut $\Sigma$. 

The $C_0$-vacuum subtracted modular energy of the state $|\Psi\>$ restricted to $\ci^+_>$ is  
\be K_0=-\tr \rho \ln \sigma_0+\tr\sigma_0\ln \sigma_0.\ee
This expression  vanishes for $\rho=\sigma_0$, as does $ A^\Sigma_0 $ when the physical geometry is the $C_0$ vacuum. Hence the constant is fixed so that 
\be K_0= -{  A^\Sigma_0 \over 4 \hbar G}.\ee

 We further define the regulated entanglement entropy
\be  S^{\rm ent}_0=-\tr  \rho\ln \rho +\tr  \sigma_0\ln \sigma_0 \ee
and the relative entropy
\be S(\rho| \sigma_0)=\tr  \rho \ln \rho-\tr  \rho\ln \sigma_0.\ee
Evidently
\be S(\rho| \sigma_0)=K_0- S^{\rm ent}_0.\ee
Positivity of relative entropy and the conjecture (\ref{cnj}) then implies the bound
\be\label{XDV} -{  A^\Sigma_0 \over 4\hbar G}\ge  S^{\rm ent}_0. \ee
We note that the renormalized area $A^\Sigma_0$ is typically negative while the entanglement entropy is typically positive. If the renormalized area and entanglement entropy both tend to zero when the cut $\Sigma$ is taken to ${\cal I}_+^+$, then it follows from (\ref{XDV}) that the change (final minus initial) $\Delta A$ in the renormalized area and the change $\Delta S^{\rm ent}$ in the entanglement entropy obey the `second law of \ip'\footnote{This highlights the differences with
the situations typically considered in  \cite{Bousso:1999xy} involving  area decreases and positive entropy fluxes. }
\be \frac{\Delta A}{4\hbar G}+\Delta S^{\rm ent} \ge 0.\ee
In this inequality, $\Delta A$ is typically positive while $\Delta S$ is typically negative, reflecting the fact that the outgoing flux  after the cut  $\Sigma$ is correlated with the flux prior to $\Sigma$ if it is to restore quantum purity.

\section*{Acknowledgements}
We are  grateful  to Daniel Harlow, J. Maldacena and Malcolm Perry for discussions, and especially to Raphael Bousso for 
discussions and for collaboration at an early stage. This work was supported in part by NSF  grant 1205550 and the Fundamental Laws Initiative at Harvard.


\begin{thebibliography}{99}


\bibitem{Bekenstein:1972tm} 
  J.~D.~Bekenstein,
  ``Black holes and the second law,''
  Lett.\ Nuovo Cim.\  {\bf 4}, 737 (1972).
  

  
\bibitem{Hawking:1974sw} 
  S.~W.~Hawking,
  ``Particle Creation by Black Holes,''
  Commun.\ Math.\ Phys.\  {\bf 43}, 199 (1975)
  Erratum: [Commun.\ Math.\ Phys.\  {\bf 46}, 206 (1976)].
  

\bibitem{Unruh:1976db} 
  W.~G.~Unruh,
  ``Notes on black hole evaporation,''
  Phys.\ Rev.\ D {\bf 14}, 870 (1976).


\bibitem{Gibbons:1977mu} 
  G.~W.~Gibbons and S.~W.~Hawking,
  ``Cosmological Event Horizons, Thermodynamics, and Particle Creation,''
  Phys.\ Rev.\ D {\bf 15}, 2738 (1977).
  
\bibitem{Gibbons:1976ue} 
  G.~W.~Gibbons and S.~W.~Hawking,
  ``Action Integrals and Partition Functions in Quantum Gravity,''
  Phys.\ Rev.\ D {\bf 15}, 2752 (1977).
  
  
 
\bibitem{Wald:1993nt} 
  R.~M.~Wald,
  ``Black hole entropy is the Noether charge,''
  Phys.\ Rev.\ D {\bf 48}, 3427 (1993)
  [gr-qc/9307038].
  
\bibitem{Iyer:1994ys} 
  V.~Iyer and R.~M.~Wald,
  ``Some properties of Noether charge and a proposal for dynamical black hole entropy,''
  Phys.\ Rev.\ D {\bf 50}, 846 (1994)
  [gr-qc/9403028].
  
  
\bibitem{'tHooft:1993gx} 
  G.~'t Hooft,
  ``Dimensional reduction in quantum gravity,''
  Salamfest 1993:0284-296
  [gr-qc/9310026].
\bibitem{Fiola:1994ir} 
  T.~M.~Fiola, J.~Preskill, A.~Strominger and S.~P.~Trivedi,
  ``Black hole thermodynamics and information loss in two-dimensions,''
  Phys.\ Rev.\ D {\bf 50}, 3987 (1994)
  [hep-th/9403137].

\bibitem{Susskind:1994vu} 
  L.~Susskind,
  ``The World as a hologram,''
  J.\ Math.\ Phys.\  {\bf 36}, 6377 (1995)
  [hep-th/9409089].
\bibitem{Jacobson:1995ab} 
  T.~Jacobson,
  ``Thermodynamics of space-time: The Einstein equation of state,''
  Phys.\ Rev.\ Lett.\  {\bf 75}, 1260 (1995)
  [gr-qc/9504004].


\bibitem{Strominger:1996sh} 
  A.~Strominger and C.~Vafa,
  ``Microscopic origin of the Bekenstein-Hawking entropy,''
  Phys.\ Lett.\ B {\bf 379}, 99 (1996)
  [hep-th/9601029].

\bibitem{Maldacena:1997re} 
  J.~M.~Maldacena,
  ``The Large N limit of superconformal field theories and supergravity,''
  Int.\ J.\ Theor.\ Phys.\  {\bf 38}, 1113 (1999)
  [Adv.\ Theor.\ Math.\ Phys.\  {\bf 2}, 231 (1998)]
  [hep-th/9711200].

\bibitem{Witten:1998qj} 
  E.~Witten,
  ``Anti-de Sitter space and holography,''
  Adv.\ Theor.\ Math.\ Phys.\  {\bf 2}, 253 (1998)
  [hep-th/9802150].
  
  
\bibitem{Susskind:1998dq} 
  L.~Susskind and E.~Witten,
  ``The Holographic bound in anti-de Sitter space,''
  hep-th/9805114.
  
  
\bibitem{Gubser:1998bc} 
  S.~S.~Gubser, I.~R.~Klebanov and A.~M.~Polyakov,
  ``Gauge theory correlators from noncritical string theory,''
  Phys.\ Lett.\ B {\bf 428}, 105 (1998)
  [hep-th/9802109].

\bibitem{Hawking:2000da} 
  S.~Hawking, J.~M.~Maldacena and A.~Strominger,
  ``de Sitter entropy, quantum entanglement and AdS / CFT,''
  JHEP {\bf 0105}, 001 (2001)
  [hep-th/0002145].
  
\bibitem{Fischler:1998st} 
  W.~Fischler and L.~Susskind,
  ``Holography and cosmology,''
  hep-th/9806039.
\bibitem{Strominger:2003br} 
  A.~Strominger and D.~M.~Thompson,
  ``A Quantum Bousso bound,''
  Phys.\ Rev.\ D {\bf 70}, 044007 (2004)
  [hep-th/0303067].
\bibitem{Ryu:2006bv} 
  S.~Ryu and T.~Takayanagi,
  ``Holographic derivation of entanglement entropy from AdS/CFT,''
  Phys.\ Rev.\ Lett.\  {\bf 96}, 181602 (2006)
  [hep-th/0603001].

\bibitem{Ryu:2006ef} 
  S.~Ryu and T.~Takayanagi,
  ``Aspects of Holographic Entanglement Entropy,''
  JHEP {\bf 0608}, 045 (2006)
  [hep-th/0605073].
  
\bibitem{Casini:2008cr} 
  H.~Casini,
  ``Relative entropy and the Bekenstein bound,''
  Class.\ Quant.\ Grav.\  {\bf 25}, 205021 (2008)
  [arXiv:0804.2182 [hep-th]].








\bibitem{Bousso:1999xy} 
  R.~Bousso,
  ``A Covariant entropy conjecture,''
  JHEP {\bf 9907}, 004 (1999)
  [hep-th/9905177].

\bibitem{Bousso:1999cb} 
  R.~Bousso,
  ``Holography in general space-times,''
  JHEP {\bf 9906}, 028 (1999)
  [hep-th/9906022].
  
  
  
  
\bibitem{Bousso:2002ju} 
  R.~Bousso,
  ``The Holographic principle,''
  Rev.\ Mod.\ Phys.\  {\bf 74}, 825 (2002)
  [hep-th/0203101].

\bibitem{Flanagan:1999jp} 
  E.~E.~Flanagan, D.~Marolf and R.~M.~Wald,
  ``Proof of classical versions of the Bousso entropy bound and of the generalized second law,''
  Phys.\ Rev.\ D {\bf 62}, 084035 (2000)
  [hep-th/9908070].



\bibitem{Bousso:2014sda} 
  R.~Bousso, H.~Casini, Z.~Fisher and J.~Maldacena,
  ``Proof of a Quantum Bousso Bound,''
  Phys.\ Rev.\ D {\bf 90}, no. 4, 044002 (2014)
  [arXiv:1404.5635 [hep-th]].

\bibitem{Bousso:2014uxa} 
  R.~Bousso, H.~Casini, Z.~Fisher and J.~Maldacena,
  ``Entropy on a null surface for interacting quantum field theories and the Bousso bound,''
  Phys.\ Rev.\ D {\bf 91}, no. 8, 084030 (2015)
  [arXiv:1406.4545 [hep-th]].


\bibitem{Bousso:2015mna} 
  R.~Bousso, Z.~Fisher, S.~Leichenauer and A.~C.~Wall,
  ``A Quantum Focussing Conjecture,''
  arXiv:1506.02669 [hep-th].


\bibitem{Strominger:2013jfa} 
  A.~Strominger,
  ``On BMS Invariance of Gravitational Scattering,''
  JHEP {\bf 1407}, 152 (2014)
  [arXiv:1312.2229 [hep-th]].

\bibitem{ash}
 A.~Ashtekar,
  ``Asymptotic Quantization of the Gravitational Field,''
  Phys.\ Rev.\ Lett.\  {\bf 46}, 573 (1981).
  \bibitem{ashb}A.~Ashtekar, ``Asymptotic Quantization: Based On 1984 Naples Lectures,''
  NAPLES, ITALY: BIBLIOPOLIS (1987) 107 P. (MONOGRAPHS AND TEXTBOOKS IN PHYSICAL SCIENCE, 2)
  
  \bibitem{Balachandran:2014hra} 
  A.~P.~Balachandran, S.~Kurkcuoglu, A.~R.~de Queiroz and S.~Vaidya,
  ``Spontaneous Lorentz Violation: The Case of Infrared QED,''
  Eur.\ Phys.\ J.\ C {\bf 75}, no. 2, 89 (2015)
  [arXiv:1406.5845 [hep-th]].



\bibitem{Casini:2013rba} 
  H.~Casini, M.~Huerta and J.~A.~Rosabal,
  ``Remarks on entanglement entropy for gauge fields,''
  Phys.\ Rev.\ D {\bf 89}, no. 8, 085012 (2014)
  [arXiv:1312.1183 [hep-th]].
  


\bibitem{Donnelly:2014fua} 
  W.~Donnelly and A.~C.~Wall,
  ``Entanglement entropy of electromagnetic edge modes,''
  Phys.\ Rev.\ Lett.\  {\bf 114}, no. 11, 111603 (2015)
  [arXiv:1412.1895 [hep-th]].


\bibitem{Donnelly:2015hxa}
  W.~Donnelly and A.~C.~Wall,
  ``Geometric entropy and edge modes of the electromagnetic field,''
  arXiv:1506.05792 [hep-th].






\bibitem{Page:1993wv} 
  D.~N.~Page,
  ``Information in black hole radiation,''
  Phys.\ Rev.\ Lett.\  {\bf 71}, 3743 (1993)
  [hep-th/9306083].

\bibitem{Page:2013dx} 
  D.~N.~Page,
  ``Time Dependence of Hawking Radiation Entropy,''
  JCAP {\bf 1309}, 028 (2013)
  [arXiv:1301.4995 [hep-th]].



\bibitem{Hawking:2016msc} 
  S.~W.~Hawking, M.~J.~Perry and A.~Strominger,
  ``Soft Hair on Black Holes,''
  arXiv:1601.00921 [hep-th].
  
  
  
  
  
\bibitem{Strominger:2014pwa} 
  A.~Strominger and A.~Zhiboedov,
  ``Gravitational Memory, BMS Supertranslations and Soft Theorems,''
  JHEP {\bf 1601}, 086 (2016)
  [arXiv:1411.5745 [hep-th]].















\bibitem{Bondi:1962px} 
  H.~Bondi, M.~G.~J.~van der Burg and A.~W.~K.~Metzner,
  ``Gravitational waves in general relativity. 7. Waves from axisymmetric isolated systems,''
  Proc.\ Roy.\ Soc.\ Lond.\ A {\bf 269}, 21 (1962).

\bibitem{Sachs:1962wk} 
  R.~K.~Sachs,
  ``Gravitational waves in general relativity. 8. Waves in asymptotically flat space-times,''
  Proc.\ Roy.\ Soc.\ Lond.\ A {\bf 270}, 103 (1962).

\bibitem{Sachs:1962zza} 
  R.~Sachs,
  ``Asymptotic symmetries in gravitational theory,''
  Phys.\ Rev.\  {\bf 128}, 2851 (1962).



\bibitem{Ashtekar:2014zsa} 
  A.~Ashtekar,
  ``Geometry and Physics of Null Infinity,''
  arXiv:1409.1800 [gr-qc].









\bibitem{Barnich:2010eb} 
  G.~Barnich and C.~Troessaert,
  ``Aspects of the BMS/CFT correspondence,''
  JHEP {\bf 1005}, 062 (2010)
  [arXiv:1001.1541 [hep-th]].

\bibitem{Barnich:2011mi} 
  G.~Barnich and C.~Troessaert,
  ``BMS charge algebra,''
  JHEP {\bf 1112}, 105 (2011)
  [arXiv:1106.0213 [hep-th]].








\bibitem{Wall:2010cj} 
  A.~C.~Wall,
  ``A Proof of the generalized second law for rapidly-evolving Rindler horizons,''
  Phys.\ Rev.\ D {\bf 82}, 124019 (2010)
  [arXiv:1007.1493 [gr-qc]].

\bibitem{Wall:2011hj} 
  A.~C.~Wall,
  ``A proof of the generalized second law for rapidly changing fields and arbitrary horizon slices,''
  Phys.\ Rev.\ D {\bf 85}, 104049 (2012)
  Erratum: [Phys.\ Rev.\ D {\bf 87}, no. 6, 069904 (2013)]
  [arXiv:1105.3445 [gr-qc]].








\end{thebibliography}
\end{document}